# Patterns of Medical Care Cost by Service Type Associated with Lung Cancer Screening

Kris Wain, PhD[1],

Mahesh Maiyani, MBA[1],

Nikki M. Carroll, MS[1],

Rafael Meza, PhD[2],

Robert T. Greenlee, PhD, MPH[3],

Christine Neslund-Dudas, PhD[4],

Michelle R. Odelberg, MPH[1],

Caryn Oshiro, PhD[5],

Debra P. Ritzwoller, PhD[1]

[1]Institute for Health Research, Kaiser Permanente Colorado, Aurora, CO

[2]Department of Integrative Oncology, BC Cancer Research Institute, Vancouver, BC, Canada

[3]Marshfield Clinic Research Institute, Marshfield Clinic Health System, Marshfield, Wisconsin

[4]Henry Ford Health System and Henry Ford Cancer Institute, Detroit, Michigan

[5]Center for Integrated Healthcare Research, Kaiser Permanente Hawaii, Honolulu, Hawaii

**ABSTRACT**

**Introduction:**  Lung cancer screening (LCS) increases early-stage cancer detection which may reduce cancer treatment costs. Little is known about how receipt of LCS affects healthcare costs in real-world clinical settings.

**Methods:**  This retrospective study analyzed utilization and cost data from the Population-based Research to Optimize the Screening Process Lung Consortium. We included individuals who met age and smoking LCS eligibility criteria and were engaged within four healthcare systems between February 5, 2015, and December 31, 2021. Generalized linear models estimated healthcare costs from the payer perspective during 12-months prior and 12-months post baseline LCS. We compared these costs to eligible individuals who did not receive LCS. Sensitivity analyses expanded our sample to age-eligible individuals with any smoking history noted in the electronic health record. Secondary analyses examined costs among a sample diagnosed with lung cancer. We reported mean predicted costs with average values for all other explanatory variables.

**Results**:  We identified 10,049 eligible individuals who received baseline LCS and 15,233 who did not receive baseline LCS. Receipt of baseline LCS was associated with additional costs of $3,698 compared to individuals not receiving LCS. Secondary analyses showed suggestive evidence that LCS prior to cancer diagnosis decreased healthcare costs compared to cancer diagnosed without screening.

**Conclusion:**  These findings suggest LCS increases healthcare costs in the year following screening. However, LCS also improves early-stage cancer detection and may reduce treatment costs following diagnosis. These results can inform future simulation models to guide LCS recommendations, and aid health policy decision makers on resource allocation.

**Introduction**

Lung cancer screening (LCS) is an effective method for early-stage cancer detection, and may reduce lung cancer treatment costs following a lung cancer diagnosis.[1-3] Current estimates predict 230,000 cases of lung cancer are diagnosed annually in the United States, with annual treatment costs approaching $20 billion USD.[4,5] In 2011, the National Lung Screening Trial (NLST) found that LCS via low-dose computed tomography reduced lung cancer mortality by 20%. Results from the NLST prompted the United States Preventive Services Task Force (USPSTF) to recommend LCS with a "B" grade in 2013 (expanded to a broader population in 2021).[6,7] In 2015, the Centers for Medicare and Medicaid Services (CMS) required most commercial insurers to exempt LCS from patient cost-sharing to promote uptake.[8] Despite declining smoking rates, the expanded 2021 USPSTF recommendations are anticipated to increase LCS uptake and improve lung cancer detection.[9] Coupled with the high cost for new cancer therapies, projections indicate lung cancer will account for more than 15% of all cancer costs.[10] Understanding LCS implementation and delivery costs is important for healthcare system decision makers, and real-world cost estimates are needed to inform future eligibility guidelines through simulation modeling.

Prior research estimating costs associated with LCS has often relied upon large randomized clinical trials (RCTs). The NLST and the Dutch-Belgian Lung Cancer Screening Trial (NELSON) estimated an incremental cost of $1,600 per individual receiving LCS.[11,12] Estimates from The Cancer Intervention and Surveillance Modeling Network (CISNET) models place the incremental cost of LCS between $870 to $980 per

individual.[13] Cost estimates from LCS trials and simulation models are typically limited to a narrow set of utilization directly related to lung cancer screening, such as diagnostic workups for positive screens and lung cancer treatment, and do not examine indirect cost increases from incidental or other findings following LCS.[14,15] Growing evidence suggests healthcare utilization following LCS in RCTs differs from utilization patterns in real-world clinical settings due to differences in the underlying population receiving baseline screening (e.g. differences in age, smoking status, comorbidity profile, etc.).[16-24] Moreover, studies leveraging claims data often lack detailed information on smoking history and are unable to accurately identify individuals eligible for LCS that do not participate in screening.[25]

There is limited information quantifying the effect LCS has on healthcare costs. To our knowledge, only three studies have examined healthcare costs associated with LCS in real-world clinical settings.[26-28] One study analyzed a small cohort of individuals residing in the Canadian province of Alberta and found no cost difference between screened and unscreened individuals.[26] Two other studies estimated LCS costs were between $273 and $740, however neither of these studies was able to identify a comparison group eligible for LCS but did not receive screening.[27,28] Our study aims to quantify changes in healthcare costs during the 12-months prior to preventive LCS relative to the 12-months post preventive LCS within a large and diverse population of eligible individuals, thus enhancing existing estimates.

**Methods**

***Study Setting and Participants***

This study was conducted within the Population-based Research to Optimize the Screening Process (PROSPR) Lung Consortium.[24] The PROSPR-Lung Consortium developed a data infrastructure for patients from five diverse healthcare systems between January 1, 2010 and December 31, 2021. Study protocol and human subjects considerations were reviewed and approved by the Kaiser Permanente Colorado Institutional Review Board.

Data from four PROSPR-Lung healthcare systems were included in our analyses: the Kaiser Permanente regions of Colorado and Hawaii, Henry Ford Health, and Marshfield Clinic Research Foundation. Our cohort included individuals who met 2013 USPSTF eligibility: between the ages of 55 and 80 with a 30 pack-year smoking history who currently smoked or quit within 15 years.[29] All individuals were required to have 12 months of engagement with their healthcare system pre- and post-index date, providing all individuals an equivalent time interval to estimate healthcare costs. We excluded individuals with a diagnosis of lung cancer prior to their index date and those who died during the 12-month follow up period.

**Treatment, Outcome and Explanatory Variables**

Individuals were categorized into two groups: those who received baseline LCS and those who did not receive baseline LCS. Baseline LCS was identified through standardized procedural codes from electronic health record (EHR) based radiology reporting. The baseline LCS date was considered the index date for those who received screening. Our comparison group not receiving LCS did not have a natural time interval during which to measure healthcare costs. To provide an equivalent time interval, the comparison group was assigned a random index date that followed the same distribution of dates as those who received LCS (Supplemental Figure SF1).[30,31]

Our primary outcome was healthcare costs from the payer perspective over a two-year time horizon. Costs were estimated using the Standardized Relative Resource Cost Algorithm (SRRCA) which assigned uniform costs to healthcare utilization consistently across time and setting.[31-33] Utilization was bucketed into components of medical care including: inpatient admissions, ambulatory care, medication dispensings, systemic therapy (oral/infused/injected agents, some of which may have been for non-cancer treatments), and a category containing all other costs (laboratory tests, virtual care, skilled nursing facility stays, home health, hospice, dialysis, and durable medical equipment). A small portion of medication dispensings were assigned abnormally high prices, these values were replaced with the median price for that medication. All other outliers were capped at the 99[th] percentile. Average costs were estimated in 2020 US dollars for the 12-month pre-index period, and the 12-month post-index period. The month of index date was used as a washout and excluded from our analyses.

Baseline explanatory variables were measured during the 12-month period prior to the index date and included data on individuals' age, sex, race and ethnicity, smoking history, comorbidity burden, and socioeconomic status. Age was categorized as follows: 55 to 64 years, 65 to 74 years, and 75 years or more. We chose these groupings because individuals aged 64 and under are traditionally not eligible for Medicare, while individuals aged 75 and older may have different comorbidity profiles. Self-Reported race and ethnic categories were mutually exclusive and included White, Black, Asian, Hispanic, Unknown, and other minority races including Native Hawaiian / Pacific Islander, Native American, and Multiracial. Minority races were collapsed into one category due to small sample size. Baseline smoking status was categorized as individuals who currently smoke, individuals who formerly smoked (<=1 year), or distal quitter (> 1 year). While 2013 USPSTF LCS recommendations required a 30 pack-year smoking history, individuals reporting a 40+ pack-year history were considered at highest risk for lung cancer. A binary indicator of < 40 pack-years or >= 40 pack-years was included to account for this increased risk.[6] Comorbidity status was calculated using the Deyo adaptation of the Charlson Comorbidity score for diagnoses in the year prior to the index.[34] Chronic obstructive pulmonary disease (COPD) was excluded from our comorbidity index and included separately as a binary covariate because it is highly correlated with both LCS and cost outcomes.[35] We included the state level Yost index as our measure of socio-economic status, a weighted combination of education, income, housing, and employment.[36] The Yost index was aggregated into quintiles, the first quintile representing the most deprived and the fifth quintile most affluent. Body Mass Index (BMI) was categorized as follows: less than 25 $kg/m^2$, 25 to 35 $kg/m^2$, and greater than 35 kg/m$^2$. Year of index date was included to control for annual changes in receipt of LCS and follow-up care. Healthcare system was included as a categorical variable to account for site-level differences in LCS and other types of utilization.

### Statistical Analyses

Baseline characteristics between the LCS group and No-LCS group were compared using chi-square tests for categorical variables. Generalized Linear Models (GLM) were employed to estimate healthcare costs. We used a log link function with a gamma distribution to account for asymmetrical and right-skewed costs. A Modified Park Test was used to evaluate our selection of link function and family distribution. A portion of our sample did not consume healthcare for all categories, resulting in $0 costs. Because the log-transformation of $0 creates undefined values, we added $1 to all cost categories to allow for consistent inclusion of all eligible individuals in our models.

Mean costs and 95% confidence intervals (CI) were estimated by utilization category in the 12-months pre- and post-index date. To account for differences in baseline characteristics by receipt of baseline LCS, costs were estimated within-group. We calculated adjusted cost ratios (CR) and 95% CIs to estimate the percentage change in healthcare costs between the pre- and post-index periods by receipt of LCS. Confidence Intervals for CRs were estimated with bootstrapping, using a random sample of 10,000 observations with replacement and 1,000 replications.

*Sensitivity Analyses*

Our primary sample included individuals eligible for preventive LCS under the 2013 USPSTF guidelines which required a 30 pack-year smoking history (fully eligible). However, pack-year history was missing for 57% of our sample meeting LCS age eligibility and is often systematically unavailable in claims or survey-based datasets. To explore the sensitivity of our results to less comprehensive smoking history data, we re-estimated our GLMs among age eligible individuals who were documented as having ever smoked cigarettes (expanded eligible).

*Secondary Analyses*

Among the subsample of our population diagnosed with lung cancer, we examined the distribution of cancer stage by receipt of baseline LCS (early-stage (American Joint Committee on Cancer (AJCC) I or II) vs late-stage (AJCC III or IV). To explore how receipt of baseline LCS prior to a diagnosis of lung cancer can affect healthcare costs, we re-estimated our GLMs for the subset of patients diagnosed with lung cancer in the 12-months post-index date. In addition, we explored methodologies to aid in understanding the causal effect LCS has on healthcare costs, including difference-in-differences (Supplemental Figures SF2 and SF3). However, we were unable to satisfy the assumptions necessary to employ difference-in-differences methodologies.

All analyses were performed using SAS® Software version 9.4 (SAS Institute Inc., Cary, North Carolina) and STATA® Statistical Software version 18.0 (StataCorp. 2019. Stata Statistical Software: Release 16. College Station, TX: StataCorp LLC).

**Results**

After applying exclusion criteria, our primary cohort consisted of 25,282 individuals

(Figure 1): 10,049 of whom received baseline LCS (39.7%; Table 1). Individuals aged

74 and older were less likely to receive LCS than those 73 and younger (6.9% vs.

20.3%). Individuals who currently smoke were more likely to receive LCS than individuals who formerly smoked (54.6% vs 42.8%). Individuals with no diagnosed chronic conditions had lower rates of LCS (31.8% vs 38.6%), while those diagnosed with COPD were more likely to receive LCS (28.2% vs 21.0%). Baseline characteristics for our expanded eligible sample are available in Supplemental Table ST1.

*Unadjusted Costs*

Unadjusted costs in our fully eligible sample showed the LCS group increased 12 month

pre- to post-index total costs by $3,614, while the No-LCS decreased pre- to post-index

total costs by $18 (Supplemental Table ST2). The largest pre- to post-index cost

differences by receipt of baseline LCS were among inpatient admissions ($1,400 LCS

vs -$691 No-LCS) and medication dispensings ($871 LCS vs $392 No-LCS).

*Pre-Index Costs (Adjusted)*

In the 12-month pre-index period, our adjusted GLM estimated the LCS group had total

costs of $17,560 (95% CI: $16,690 to $18,477; Figure 2), which was not significantly

different than total costs among the No-LCS group $16,447 (95% CI: $15,715 to

$17,213). Stratified by category of utilization, the LCS group had lower systemic therapy

costs (*LCS* $85, 95% CI: $78 to $93; *No-LCS* $117 CI: $108 to $127; Figure 3). We did not observe significant pre-index cost differences by receipt of LCS for inpatient admissions, ambulatory visits, medication dispensing, or other costs. Sensitivity analyses that examined costs among our expanded eligible sample showed individuals who received LCS had total costs of $17,810 (95% CI: $17,256 to $18,381), significantly more than total costs among the No-LCS group of $15,725 (95% CI: $15,441 to $16,014; Figure 2).

*Post-Index Costs (Adjusted)*

In the 12-month post-index period, our adjusted GLM estimated total costs were significantly higher among the LCS group compared to the No-LCS group (*LCS* $22,104, 95% CI: $20,935 to $23,337; *No-LCS* $17,292, 95% CI: $16,476 to $18,152; Figure 2). The LCS group had higher costs for inpatient admissions (*LCS* $8,980, 95% CI: $8,052 to $10,014; *No-LCS* $7,198, 95% CI: $6,536 to $7,928), ambulatory care (*LCS* $3,731, 95% CI: $3,548 to $3,923; *No-LCS* $2,786, 95% CI: $2,664 to $2,913), medication dispensing (*LCS* $5,388, 95% CI: $4,996 to $5,811; *No-LCS* $4,579, 95% CI: $4,281 to $4,899), systemic therapy (*LCS* $165, 95% CI: $150 to $181; *No-LCS* $90, 95% CI: $83 to $97), and other costs (*LCS* $1,164, 95% CI: $1,104 to $1,226; *No-LCS* $989, 95% CI: $944 to $1,036; Figure 3). Sensitivity analyses that examined costs among our expanded eligible sample estimated significantly higher total costs among the LCS group as compared to the No-LCS group (*LCS* $24,021, 95% CI: $23,233 to $24,837; *No-LCS* $17,574, 95% CI: $17,242 to $17,914; Figure 2).

*Pre-Index to Post-Index Cost Ratios*

Within our primary sample, pre- to post-index total costs increased by 27% (CR=1.27, 95% CI: 1.26 to 1.27) among the LCS group, and 5% among the No-LCS group (CR=1.05, 95% CI: 1.05 to 1.06; Table 2). Cost increases were larger in the LCS group for many utilization categories including inpatient admissions (LCS CR=1.36, 95% CI: 1.35 to 1.37; No-LCS CR=1.11, 95% CI: 1.11 to 1.12), ambulatory care (LCS CR=1.30, 95% CI: 1.30 to 1.31; No-LCS CR=1.04, 95% CI: 1.04 to 1.05), medication dispensing (LCS CR=1.16, 95% CI: 1.16 to 1.16; No-LCS CR=1.06, 95% CI: 1.06 to 1.07), and systemic therapy (LCS CR=2.13, 95% CI: 2.07 to 2.18; No-LCS CR=0.86, 95% CI: 0.84 to 0.88). Sensitivity analyses examining our expanded eligible sample found the LCS group increased total health costs by 40%, as compared to a 14% increase in the No-LCS group (LCS CR=1.40, 95% CI: 1.39 to 1.41; No-LCS CR=1.14, 95% CI: 1.13 to 1.15).

*Distribution of Cancer Stage by Receipt of LCS*

Within our primary sample we identified 226 individuals with an incident lung cancer diagnosis in the 12-months post-index: 149 individuals that received LCS (1.5%) and 77 individuals that did not receive LCS (0.5%; Supplemental Table ST3). Within the LCS group, 58.4% of diagnoses were early-stage, while 41.6% of diagnoses were early-stage in the No-LCS group (p=0.02). We identified 486 lung cancer diagnoses in our expanded sample, of which 59.8% were early-stage in the LCS group, as compared to 36.3% early-stage in the No-LCS group (p<0.01)

*Costs Following Lung Cancer Diagnoses by Receipt of LCS*

Within our primary sample, total cost estimates among individuals diagnosed with lung cancer in the pre-index period were not statistically different between the LCS group and the No-LCS group (*LCS* $13,930, 95% CI: $8,295 to $23,394; *No-LCS* $11,596, 95% CI: $6,980 to $19,265; Supplemental Figure SF4). Post-index costs were larger in the No-LCS group, but not statistically different from the LCS group (*LCS* $61,945, 95% CI: $47,506 to $80,772; *No-LCS* $69,023, 95% CI: $52,132 to $91,388). Sensitivity analyses examining our expanded eligible sample found no statistically significant pre-index cost differences between groups (*LCS* $10,824, 95% CI: $7,642 to $15,330; *No-LCS* $9,621, 95% CI: $6,903 to $13,410). However, post-index costs were larger in the No-LCS group, but not statistically different from the LCS group (*LCS* $65,973, 95% CI: $54,307 to $80,145; *No-LCS* $83,266, 95% CI: $69,035 to $100,432).

**Discussion**

Among an LCS-eligible population of patients receiving primary and specialty care within real-world community-based healthcare systems, we found receipt of LCS increased annual healthcare costs by $3,698 more than individuals not receiving LCS. Sensitivity analyses examining an expanded population of age-eligible individuals documented as having ever smoked cigarettes found receipt of LCS increased annual healthcare costs by $4,362 more than individuals not receiving LCS. In contrast to prior studies which estimated the incremental cost of LCS between $0 and $1,600, this study found larger incremental costs associated with LCS.[11-13,26-28] While trials and simulation models typically restrict costs to utilization directly related to lung cancer, we estimate comprehensive healthcare costs during the 12 month period following LCS. Of the three prior LCS costing studies that we are aware of, one study estimated costs associated with LCS using a small homogeneous cohort (> 96% Non-Hispanic White) with universal healthcare coverage, findings from this study may not be generalizable to more diverse populations.[26,37] The second study was unable to identify a suitable comparison group of individuals eligible for LCS who did not undergo screening, thus it reported a cost per LCS episode rather than changes in costs following screening.[27] While a third study estimated the total cost of LCS in the United States based upon cross-sectional survey data from 7 states, and was not able to examine cost changes pre- vs post-LCS.[28] We estimated individuals who received LCS had larger pre- to post-index cost increases for inpatient admissions than those not receiving LCS. This is an expected finding given the higher rate of lung cancer diagnoses among individuals receiving LCS, coupled with an average of 1.6 unplanned inpatient admissions in the year following diagnosis.[38] Our findings also showed that LCS is linked to higher costs for ambulatory care and medication dispensings, which may reflect short-term cost increases related to smoking cessation efforts.[39] We did not observe clinically meaningful cost increases for systemic therapy. While novel cancer treatments can be very expensive, only 0.9% of our primary sample received a lung cancer diagnosis within the year after LCS, limiting our ability to detect changes in treatment costs.[40]

Secondary analyses found screen-detected lung cancer was more likely to be early-stage than cancer diagnosed without LCS, a finding that aligns with prior studies.[1,3,21] While our follow-up period was limited to 12 months and may not completely capture the cancer treatment episode, we observed suggestive evidence that screen-detected lung cancers had lower initial treatment costs than cancers diagnosed without LCS.[2,5,41,42] In our sample, the number of days from index date to lung cancer diagnosis was shorter among those receiving LCS (mean of 90 days from index to cancer diagnosis) than those who did not receive screening (mean of 199 days from index to cancer diagnosis). This resulted in the LCS group having longer follow-up periods during which cancer treatment is likely, potentially overstating post-index cost estimates as compared to the No-LCS group. If follow-up time after cancer diagnoses was equivalent between groups, we may have observed larger reductions in healthcare costs among those receiving LCS. Should rates of preventive LCS continue to increase, future analyses should focus on the cost differentials related to treatment and surveillance associated with increased early-stage diagnoses, which may result in long run cost savings from LCS.[3,43,44]

In addition to GLM models reported in our primary analyses, we explored a difference-in-differences (D-I-D) methodology to estimate causal effects of LCS on subsequent healthcare costs (Supplemental Materials 2). Within our primary sample, we were unable to satisfy the primary D-I-D assumption that healthcare costs between those receiving LCS and those not receiving LCS were on parallel trends prior to index date. However, within our expanded sample, pre-index healthcare costs between the LCS and No-LCS groups appeared to satisfy the parallel trends assumption. This suggests studies determining LCS eligibility using datasets with incomplete or missing smoking history should use caution when applying causal methods.[45]

This study has several limitations. Eligible individuals may have been offered the choice to accept or refuse LCS. If individuals who received LCS were increasing healthcare utilization for other reasons such as newly diagnosed comorbid conditions or enrollment in a health plan with more generous benefits, our results may overstate the costs associated with LCS.[23,46,47] We did not examine costs beyond the first year of follow-up, nor do we explore end-of-life costs which may be considerably higher.[5,42] We required our cohort to be continuously engaged with a healthcare system for 12-months pre- and post-index. Individuals who become uninsured following a positive LCS may experience delays in care, resulting in higher downstream treatment costs.[48] Our study identified systemic therapy using validated cancer-related look-up tables, however some agents may have been for non-cancer treatments and dispensed in both the pre-index and post-index periods, limiting our ability to detect precise cost increases for cancer related systemic therapy.[49,50]

In conclusion, our results suggest receipt of LCS increased healthcare costs in the year following screening. We provided descriptive evidence confirming previous findings that LCS increased early-stage diagnosis, we also showed suggestive evidence that LCS decreased treatment costs following a cancer diagnosis. We found pre- to post-index cost increases were comparable between our primary sample that fully met 2013 USPSTF eligibility and our expanded sample among age eligible individuals who were documented as having ever smoked cigarettes. In addition to providing cost inputs for future simulation modeling, our findings can assist health policy decision makers by informing on medical care costs associated with implementation of LCS and appropriate resource allocation. Future studies can examine longer time horizons and focus on cancer treatment costs, including systemic therapy, by receipt of LCS. Simulation modeling can examine whether screening can result in cost-savings if uptake significantly improves.

**Figure 1** Consort Diagram

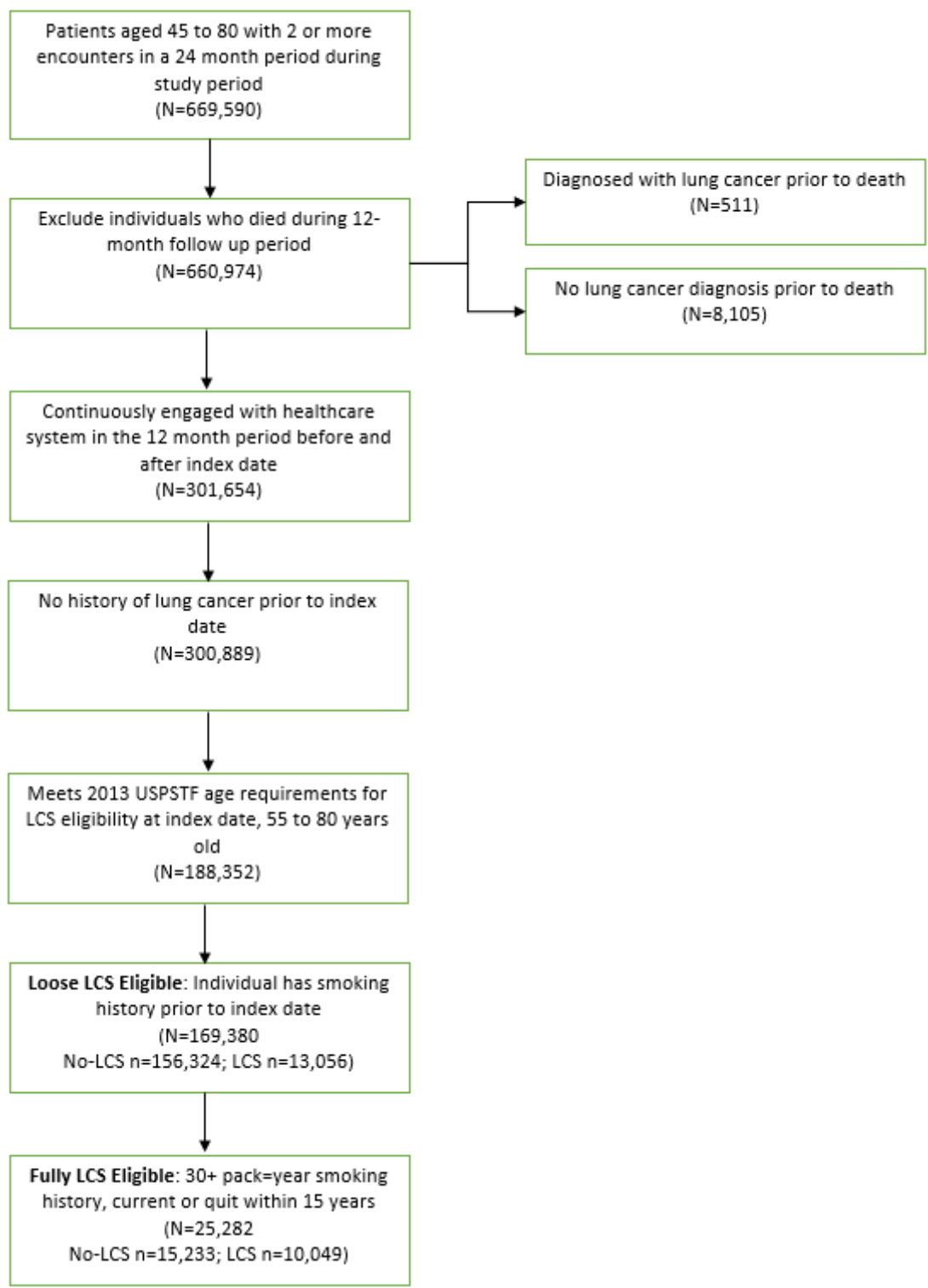

**Table 1** Demographic characteristics of individuals fully eligible for lung cancer screening by receipt of baseline lung cancer screening during the study period

| | Receipt of Baseline Lung Cancer Screening | | | |
|---|---|---|---|---|
| | *No* | *Yes* | *Total* | p-value |
| **Unique Patients, N (%)** | 15,233 (60.3) | 10,049 (39.7) | 25,282 | |
| | | | | |
| **Site, N (%)** | | | | < .01 |
| Site 1 | 7,353 (48.3) | 4,291 (42.7) | 11,644 (46.1) | |
| Site 2 | 5,008 (32.9) | 4,539 (45.2) | 9,547 (37.8) | |
| Site 3 | 1,965 (12.9) | 762 (7.6) | 2,727 (10.8) | |
| Site 4 | 907 (6.0) | 457 (4.5) | 1,364 (5.4) | |
| | | | | |
| **Age at Index Date, N (%)** | | | | < .01 |
| 55-64 | 6,092 (40.0) | 4,765 (47.4) | 10,857 (42.9) | |
| 65-74 | 6,054 (39.7) | 4,589 (45.7) | 10,643 (42.1) | |
| 74 and older | 3,087 (20.3) | 695 (6.9) | 3,782 (15.0) | |
| | | | | |
| **Gender, N (%)** | | | | 0.24 |
| Female | 6,952 (45.6) | 4,662 (46.4) | 11,614 (45.9) | |
| Male | 8,281 (54.4) | 5,387 (53.6) | 13,668 (54.1) | |
| | | | | |
| **Race/Ethnicity, N (%)** | | | | < .01 |
| Asian | 632 (4.1) | 334 (3.3) | 966 (3.8) | |
| Black | 1,496 (9.8) | 1,039 (10.3) | 2,535 (10.0) | |
| Hawaiian Pacific Islander/Native American/Mutli Race/Other | 919 (6.0) | 456 (4.5) | 1,375 (5.4) | |
| Hispanic | 656 (4.3) | 508 (5.1) | 1,164 (4.6) | |
| White | 11,011 (72.3) | 7,450 (74.1) | 18,461 (73.0) | |
| Unknown | 519 (3.4) | 262 (2.6) | 781 (3.1) | |
| | | | | |
| **Yost Index, Census based, at index date** | | | | < .01 |
| 1st quintile (lowest) | 3,132 (20.6) | 1,903 (18.9) | 5,035 (19.9) | |
| 2nd quintile | 2,987 (19.6) | 1,904 (18.9) | 4,891 (19.3) | |
| 3rd quintile | 3,642 (23.9) | 2,352 (23.4) | 5,994 (23.7) | |
| 4th quintile | 2,919 (19.2) | 2,083 (20.7) | 5,002 (19.8) | |
| 5th quintile (highest) | 2,553 (16.8) | 1,807 (18.0) | 4,360 (17.2) | |
| **Baseline Smoking Status** | | | | < .01 |
| Current | 6,516 (42.8) | 5,484 (54.6) | 12,000 (47.5) | |
| Former | 8,717 (57.2) | 4,565 (45.4) | 13,282 (52.5) | |
| | | | | |
| **Time Since Quit Smoking** | | | | < .01 |
| Current Smoker | 6,516 (42.8) | 5,484 (54.6) | 12,000 (47.5) | |
| Less Than 1 Year | 2,401 (15.8) | 877 (8.7) | 3,278 (13.0) | |
| 1 Year or more | 6,316 (41.5) | 3,688 (36.7) | 10,004 (39.6) | |
| | | | | |
| **Pack Years Smoked** | | | | < .01 |
| Less than 40 Years | 5,243 (34.4) | 2,846 (28.3) | 8,089 (32.0) | |
| More than 40 Years | 9,990 (65.6) | 7,203 (71.7) | 17,193 (68.0) | |
| | | | | |
| **Charlson Comorbidity Index** | | | | < .01 |
| 0 | 5,879 (38.6) | 3,195 (31.8) | 9,074 (35.9) | |
| 1 | 3,202 (21.0) | 2,965 (29.5) | 6,167 (24.4) | |
| 2 | 2,094 (13.7) | 1,623 (16.2) | 3,717 (14.7) | |
| 3+ | 4,058 (26.6) | 2,266 (22.5) | 6,324 (25.0) | |
| | | | | |
| **BMI** | | | | < .01 |
| Less than 25 | 4,416 (29.0) | 3,053 (30.4) | 7,469 (29.5) | |
| 25 to 35 | 8,314 (54.6) | 5,592 (55.6) | 13,906 (55.0) | |
| More than 35 | 2,503 (16.4) | 1,404 (14.0) | 3,907 (15.5) | |
| | | | | |
| **COPD Diagnosis, 1 year pre-index** | | | | |
| Yes | 3,192 (21.0) | 2,830 (28.2) | 6,022 (23.8) | < .01 |
| No | 12,041 (79.0) | 7,219 (71.8) | 19,260 (76.2) | |

- Sites have been masked to prevent unintended patient identification.

**Figure 2**  Adjusted GLM Estimated Total Healthcare Costs Pre vs Post Index Date by Receipt of Lung Cancer Screening

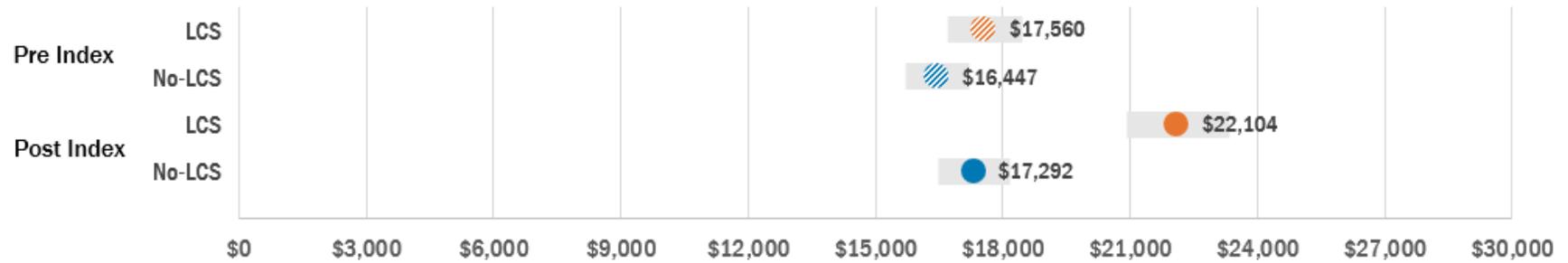

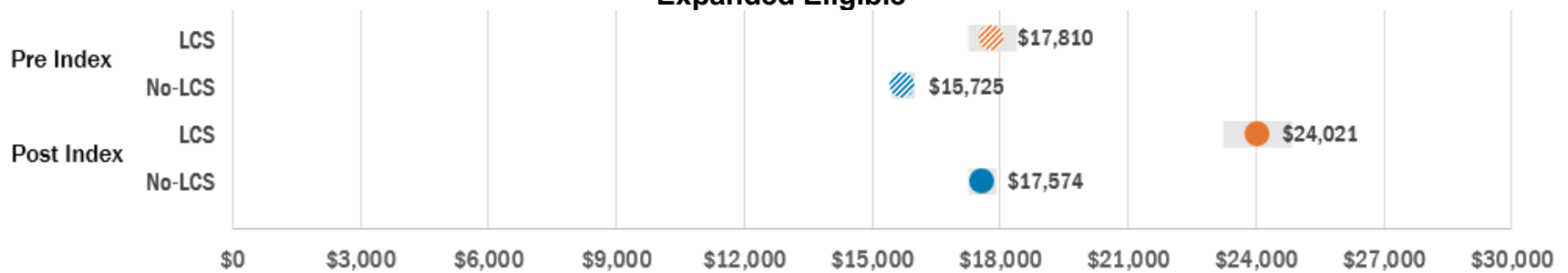

- Results reported as predicted costs with mean values for all other explanatory variables.
- Explanatory variables include age, gender, race, ethnicity, site, Yost Index, smoking status, Charlson Comorbidity Index, BMI, and diagnosis of COPD
- Fully eligible requirements are 55 to 80 years of age with a 30-pack year history and currently smoke or quit within the last 15 years.
- Loose eligibility requirements are 55 to 80 years of age with any amount of reported smoking history.
- Pre-index period includes 12 months of cost prior to month of index, post-index period includes 12 months of costs after the month of index. Month of index is excluded from results.
- other costs included: laboratory tests, virtual care, skilled nursing facility stays, home health, hospice, dialysis, and durable medical equipment.

**Figure 3**  Adjusted GLM Estimated Costs for Certain Healthcare Components Pre vs Post Index Date by Receipt of Lung Cancer Screening

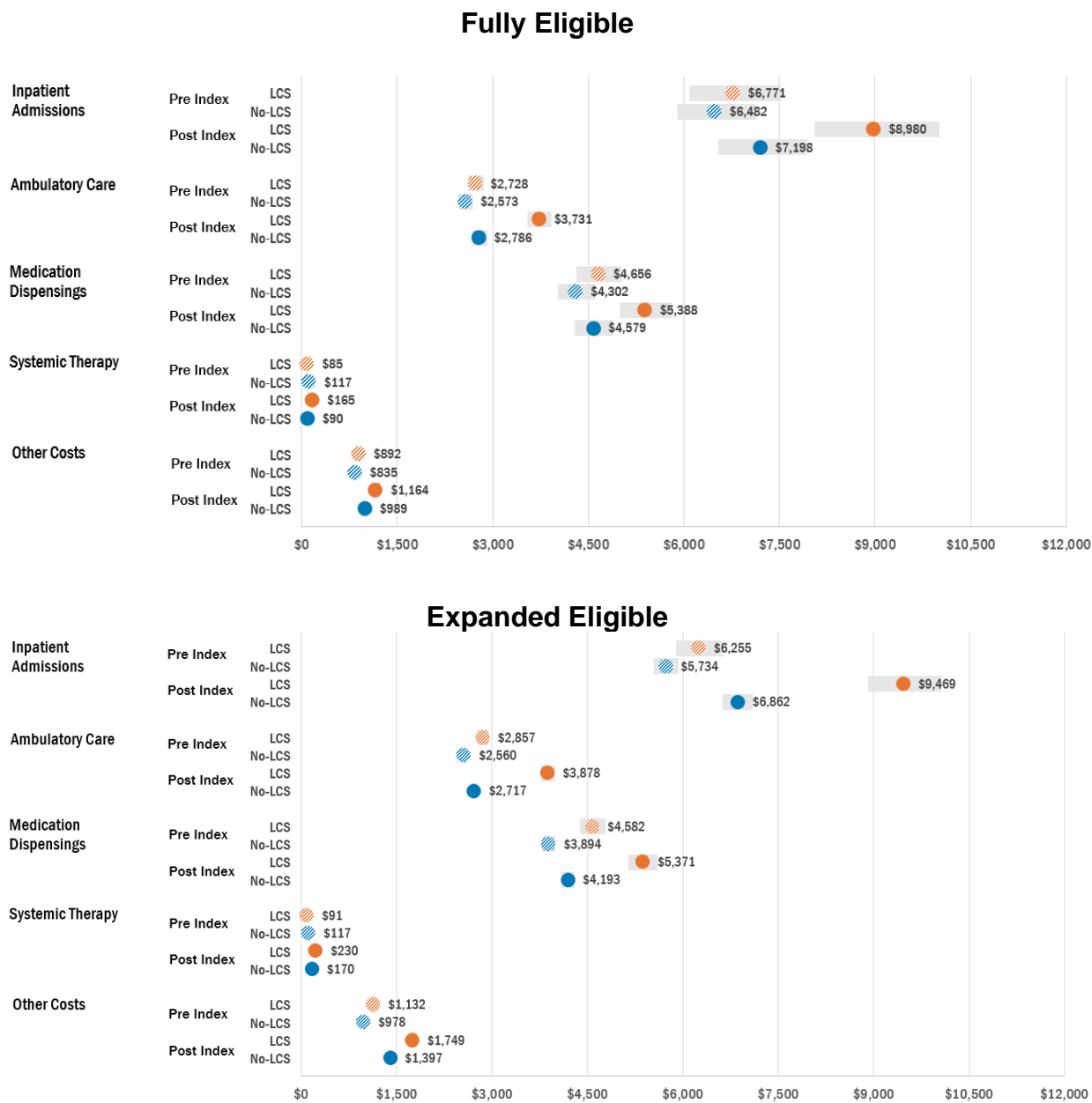

- Results reported as predicted costs with mean values for all other explanatory variables.
- Explanatory variables include age, gender, race, ethnicity, site, Yost Index, smoking status, Charlson Comorbidity Index, BMI, and diagnosis of COPD
- Fully eligible requirements are 55 to 80 years of age with a 30-pack year history and currently smoke or quit within the last 15 years.
- Loose eligibility requirements are 55 to 80 years of age with any amount of reported smoking history.
- Pre-index period includes 12 months of cost prior to month of index, post-index period includes 12 months of costs after the month of index. Month of index is excluded from results.
- other costs included: laboratory tests, virtual care, skilled nursing facility stays, home health, hospice, dialysis, and durable medical equipment.

**Table 2**  Adjusted Cost Ratio of Post-index/Pre-index Mean Healthcare Costs by Receipt of Lung Cancer Screening

**Fully Eligible**

| Medical Care Component | Receipt of Baseline Lung Cancer Screening | |
|---|---|---|
| | No | Yes |
| Total Costs | 1.05 (1.05 - 1.06) | 1.27 (1.26 - 1.27) |
| Inpatient Admissions | 1.11 (1.11 - 1.12) | 1.36 (1.35 - 1.37) |
| Ambulatory Care | 1.04 (1.04 - 1.05) | 1.30 (1.30 - 1.31) |
| Medication Dispensings | 1.06 (1.06 - 1.07) | 1.16 (1.16 - 1.16) |
| Systemic Therapy | 0.86 (0.84 – 0.88) | 2.13 (2.07 – 2.18) |
| Other Costs | 1.15 (1.14 - 1.16) | 1.24 (1.23 – 1.25) |

**Expanded Eligible**

| Medical Care Component | Receipt of Baseline Lung Cancer Screening | |
|---|---|---|
| | No | Yes |
| Total Costs | 1.14 (1.13 - 1.15) | 1.40 (1.39 - 1.41) |
| Inpatient Admissions | 1.27 (1.26 - 1.28) | 1.76 (1.75 - 1.77) |
| Ambulatory Care | 1.04 (1.03 - 1.05) | 1.30 (1.29 - 1.31) |
| Medication Dispensings | 1.09 (1.08 - 1.10) | 1.18 (1.17 - 1.19) |
| Systemic Therapy | 1.47 (1.44 - 1.50) | 3.29 (3.20 – 3.38) |
| Other Costs | 1.48 (1.47 - 1.49) | 1.59 (1.58 - 1.60) |

- Ratio of average post-index cost divided by average pre-index cost, 95% confidence intervals in parenthesis
- Costs were estimated in separate models for pre-index and post-index
- Cost estimates adjusted for same list of covariates as primary analysis
- other costs included: laboratory tests, virtual care, skilled nursing facility stays, home health, hospice, dialysis, and durable medical equipment.

**Figure SF1** Distribution of Baseline LCS Dates Among Group Receiving Lung Cancer Screening Compared to Distribution of Randomly Assigned Index Dates for Non-Screened Group

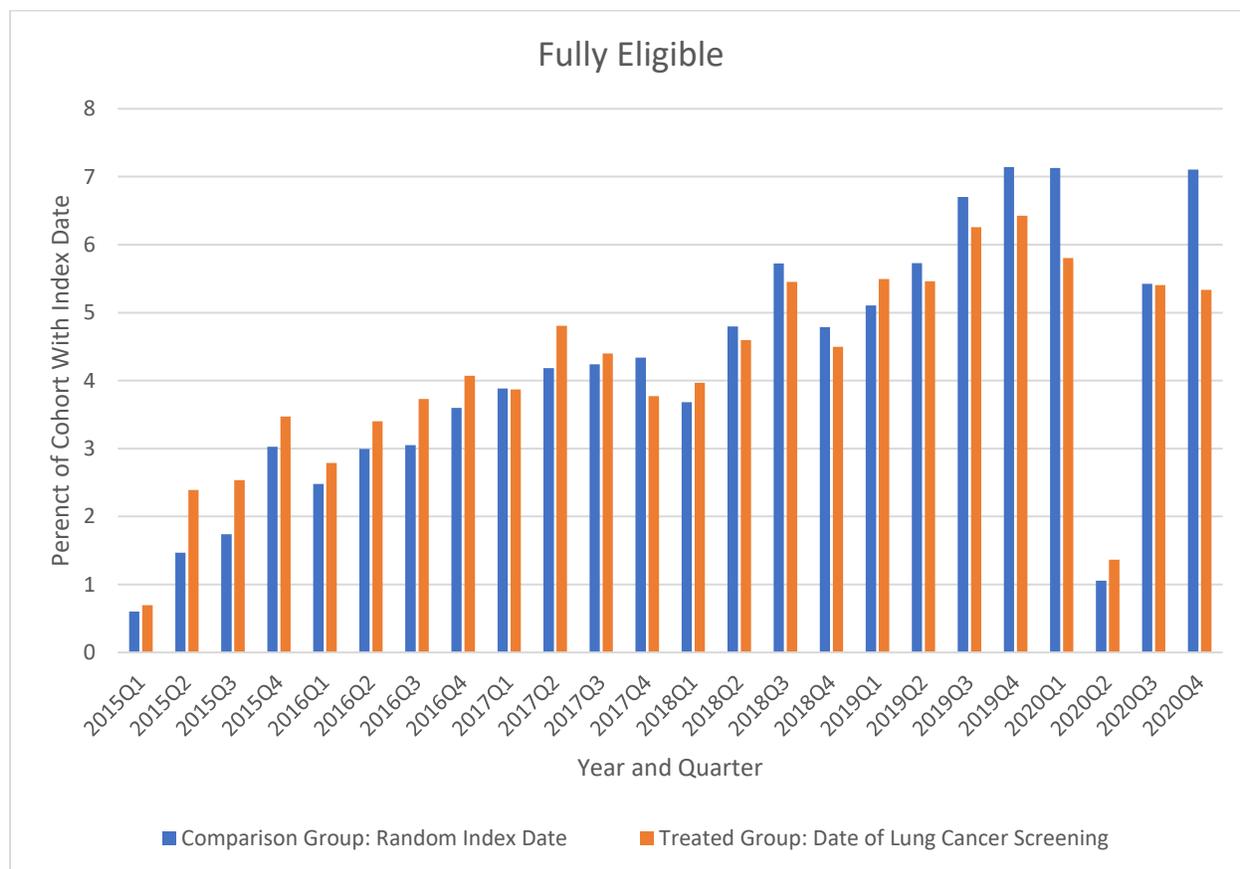

- Index dates are reported as percentage of group receiving LCS and percentage of group not receiving LCS
- Data represents sample fully eligible under 2013 USPSTF guidelines

Among individuals who received LCS, an index date was assigned on the date they received baseline LCS. Our comparison group of individuals who were not screened did not have a corresponding date that could be used to index healthcare costs. To allow for appropriate comparison between groups, a random index date was assigned to those not receiving LCS. We derived this random index date using a method similar to Jacob et al.[30] The study period was divided into quarters by year, then we mapped the distribution of baseline LCS dates for each quarter. Index dates were randomly generated for the comparison group following the distribution of dates for those that received LCS. Healthcare costs were then estimated for one year period pre- and post-random index date. Individuals in the comparison group who did not meet either age or smoking LCS eligibility requirements were excluded from the sample.

**Figure SF2** *Expanded Eligible*: Exploration of Difference in Differences Model Assumptions by Lung Cancer Screening Eligibility: Mean Unadjusted Total Costs

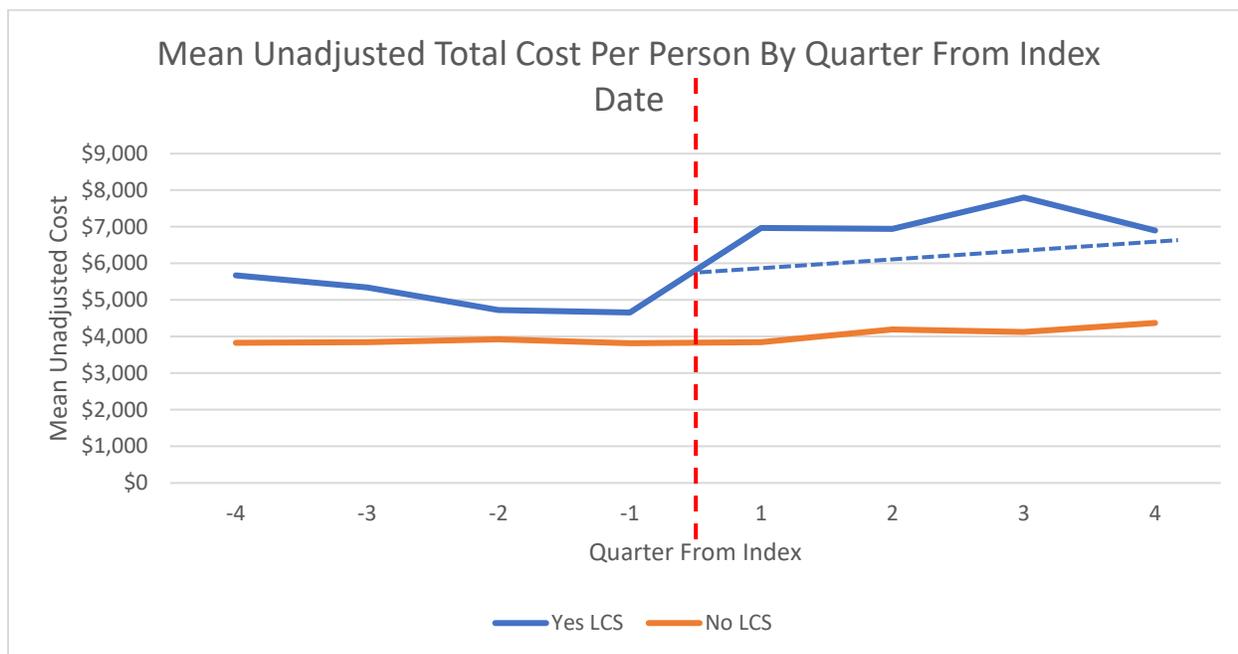

**Figure SF3** *Fully Eligible*: Exploration of Difference in Differences Model Assumptions by Lung Cancer Screening Eligibility: Mean Unadjusted Total Costs

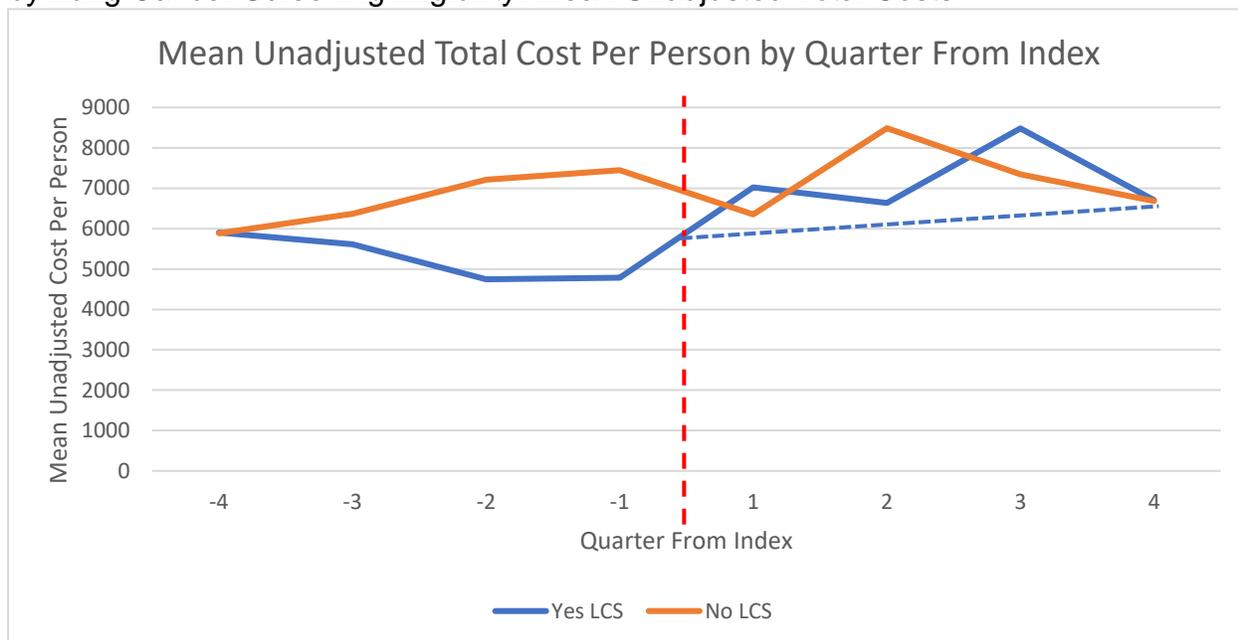

In addition to GLM models reported in our primary analysis, we explored difference in differences (D-I-D) models that can achieve causal inference when properly specified.

The primary assumption for D-I-D is that in the absence of lung cancer screening, the difference in healthcare costs between the screened and unscreened groups is constant over time. When we examined a sample of individuals who met the 2013 USPSTF age criteria (55 to 80) but did not consider smoking history, the difference in unadjusted total healthcare costs between groups appeared to be relatively constant prior to the index date (Figure SF2). However, when we limited our cohort to individuals meeting USPSTF smoking history eligibility requirements (30 pack-year history and current smoker or quit within last 15 years), the difference in unadjusted total healthcare costs between groups was not consistent prior to the index date and the parallel trends assumption is violated (Figure SF3). Researchers using claims data or survey data that lack smoking history may erroneously believe they have met the primary assumptions for D-I-D methods. Caution should be taken when determining LCS eligibility for causal inference when smoking history is missing.

**Table ST1** Expanded Eligible Sample: Demographic characteristics of individuals fully eligible for lung cancer screening by receipt of baseline lung cancer screening during the study period

| | Receipt of Baseline Lung Cancer Screening | | | |
| | No | | Yes | p-value |
| | N | % | N | |
| --- | --- | --- | --- | --- |
| **Unique Patients, N (%)** | 156,324 (92.3) | 13,056 (7.7) | 169,380 | |
| | | | | |
| **Site, N (%)** | | | | < .01 |
| HFHS | 61,612 (39.4) | 5,376 (41.2) | 66,988 (39.5) | |
| KPCO | 46,543 (29.8) | 4,817 (36.9) | 51,360 (30.3) | |
| KPHI | 19,140 (12.2) | 884 (6.8) | 20,024 (11.8) | |
| MCRF | 29,029 (18.6) | 1,979 (15.2) | 31,008 (18.3) | |
| | | | | |
| **Age at Index Date, N (%)** | | | | < .01 |
| 55-64 | 61,788 (39.5) | 6,225 (47.7) | 68,013 (40.2) | |
| 65-74 | 59,391 (38.0) | 5,943 (45.5) | 65,334 (38.6) | |
| 74 and older | 35,145 (22.5) | 888 (6.8) | 36,033 (21.3) | |
| | | | | |
| **Gender, N (%)** | | | | < .01 |
| Female | 79,232 (50.7) | 6,136 (6955 | 85,368 (50.4) | |
| Male | 77,092 (49.3) | 5,513 (6101 | 82,605 (48.8) | |
| | | | | |
| **Race/Ethnicity, N (%)** | | | | < .01 |
| White | 109,080 (69.8) | 9,853 (75.5) | 118,933 (70.2) | |
| Black | 18,522 (11.8) | 1,381 (10.6) | 19,903 (11.8) | |
| Asian | 7,506 (4.8) | 395 (3.0) | 7,901 (4.7) | |
| Hispanic | 7,913 (5.1) | 562 (4.3) | 8,475 (5.0) | |
| Hawaiian Pacific Islander/Native American/Mutli Race/Other | 8,039 (5.1) | 541 (4.1) | 8,580 (5.1) | |
| Unknown | 5,264 (3.4) | 324 (2.5) | 5,588 (3.3) | |
| | | | | |
| **Yost Index, Census based, at index date** | | | | < .01 |
| 1st quintile (lowest) | 30,444 (19.5) | 2,599 (19.9) | 33,043 (19.5) | |
| 2nd quintile | 30,491 (19.5) | 2,699 (20.7) | 33,190 (19.6) | |
| 3rd quintile | 35,333 (22.6) | 3,088 (23.7) | 38,421 (22.7) | |
| 4th quintile | 29,635 (19.0) | 2,567 (19.7) | 32,202 (19.0) | |
| 5th quintile (highest) | 30,421 (19.5) | 2,103 (16.1) | 32,524 (19.2) | |
| | | | | |
| **Baseline Smoking Status** | | | | < .01 |

| | | | |
|---|---|---|---|
| Current | 32,854 (21.0) | 7,228 (55.4) | 40,082 (23.7) |
| Former | 123,470 (79.0) | 5,828 (44.6) | 129,298 (76.3) |

| **Charlson Comorbidity Index** | | | | < .01 |
|---|---|---|---|---|
| 0 | 78,075 (49.9) | 4,241 (32.5) | 82,316 (48.6) | |
| 1 | 29,653 (19.0) | 3,857 (29.5) | 33,510 (19.8) | |
| 2 | 19,587 (12.5) | 2,133 (16.3) | 21,720 (12.8) | |
| 3+ | 29,009 (18.6) | 2,825 (21.6) | 31,834 (18.8) | |

| **BMI** | | | | < .01 |
|---|---|---|---|---|
| Less than 25 | 38,386 (24.6) | 3,917 (30.0) | 42,303 (25.0) | |
| 25 to 35 | 92,093 (58.9) | 7,257 (55.6) | 99,350 (58.7) | |
| More than 35 | 25,845 (16.5) | 1,882 (14.4) | 27,727 (16.4) | |

| **COPD Diagnosis** | 12,214 (7.8) | 3,678 (28.2) | 15,892 (9.4) | < .01 |
|---|---|---|---|---|

- Sites have been masked to prevent unintended patient identification.

**Table ST2** Unadjusted Cost Differences by Receipt of Lung Cancer Screening and Healthcare Component

**Fully Eligible**

| | | | | | | |
|---|---|---|---|---|---|---|
| | **Receipt of Baseline Lung Cancer Screening** | | | | | |
| | Yes | | | No | | |
| **Medical Care Component** | n=10,049 | | | n=15,233 | | |
| | Pre Index | Post Index | Difference | Pre Index | Post Index | Difference |
| Inpatient Admissions | $7,861 | $9,261 | $1,400 | $9,007 | $8,316 | -$691 |
| Ambulatory Care | $1,578 | $2,002 | $424 | $1,465 | $1,486 | $20 |
| Medication Dispensings | $4,750 | $5,621 | $871 | $4,396 | $4,788 | $392 |
| Systemic Therapy | $168 | $354 | $185 | $432 | $461 | $29 |
| Other Costs | $1,889 | $2,622 | $733 | $2,521 | $2,752 | $231 |
| *Total* | $16,246 | $19,860 | $3,614 | $17,822 | $17,804 | -$18 |

**Expanded Eligible**

| | | | | | | |
|---|---|---|---|---|---|---|
| | **Receipt of Baseline Lung Cancer Screening** | | | | | |
| | Yes | | | No | | |
| **Medical Care Component** | n=13,056 | | | n=156,324 | | |
| | Pre Index | Post Index | Difference | Pre Index | Post Index | Difference |
| Inpatient Admissions | $8,472 | $10,499 | $2,028 | $8,113 | $8,089 | -$24 |
| Ambulatory Care | $1,647 | $2,092 | $445 | $1,447 | $1,498 | $51 |
| Medication Dispensings | $4,557 | $5,328 | $771 | $3,553 | $3,875 | $321 |
| Systemic Therapy | $171 | $296 | $125 | $363 | $414 | $51 |
| Other Costs | $1,744 | $2,386 | $641 | $2,098 | $2,362 | $264 |
| *Total* | $16,591 | $20,602 | $4,010 | $15,574 | $16,237 | $663 |

- Fully eligible requirements are 55 to 80 years of age with a 30-pack year history and currently smoke or quit within the last 15 years.
- Expanded eligibility requirements are 55 to 80 years of age with any history of smoking.
- Pre-index period includes 12 months of cost prior to month of index, post-index period includes 12 months of costs after the month of index. Month of index is excluded from results.
- other costs included: laboratory tests, virtual care, skilled nursing facility stays, home health, hospice, dialysis, and durable medical equipment.

**Table ST3**  Stage of Lung Cancer Diagnoses by Receipt of Lung Cancer Screening

**Fully Eligible** (n=226)

| Stage at Diagnosis | Receipt of Baseline Lung Cancer Screening | | | | p-value |
| | Yes | | No | | |
| | N | % | N | % | |
|---|---|---|---|---|---|
| *I* | 87 | 58.4% | 32 | 41.6% | 0.0171 |
| *II* | 15 | 10.1% | 6 | 7.8% | |
| *III* | 26 | 17.4% | 14 | 18.2% | |
| *IV* | 13 | 8.7% | 18 | 23.4% | |
| *Missing Stage* | 8 | 5.4% | 7 | 9.1% | |
| *Total* | 149 | | 77 | | |

**Expanded Eligible** (n=486)

| Stage at Diagnosis | Receipt of Baseline Lung Cancer Screening | | | | |
| | Yes | | No | | |
| | N | % | N | % | |
|---|---|---|---|---|---|
| *I* | 116 | 59.8% | 106 | 36.3% | <.0001 |
| *II* | 23 | 11.9% | 19 | 6.5% | |
| *III* | 30 | 15.5% | 55 | 18.8% | |
| *IV* | 14 | 7.2% | 88 | 30.1% | |
| *Missing Stage* | 11 | 5.7% | 24 | 8.2% | |
| *Total* | 194 | | 292 | | |

- Lung cancer diagnoses identified through site-level tumor registries.
- Full LCS eligibly requirements are 55 to 80 years of age with a 30-pack year history and currently smoke or quit within the last 15 years.
- Pre-index period includes 12 months of cost prior to month of index, post-index period includes 12 months of costs after the month of index. Month of index is excluded from results.

**Figure SF4**  Adjusted GLM Estimated Total Healthcare Costs Pre vs Post Index Date by Receipt of Lung Cancer Screening Among Patients Diagnosed With Lung Cancer

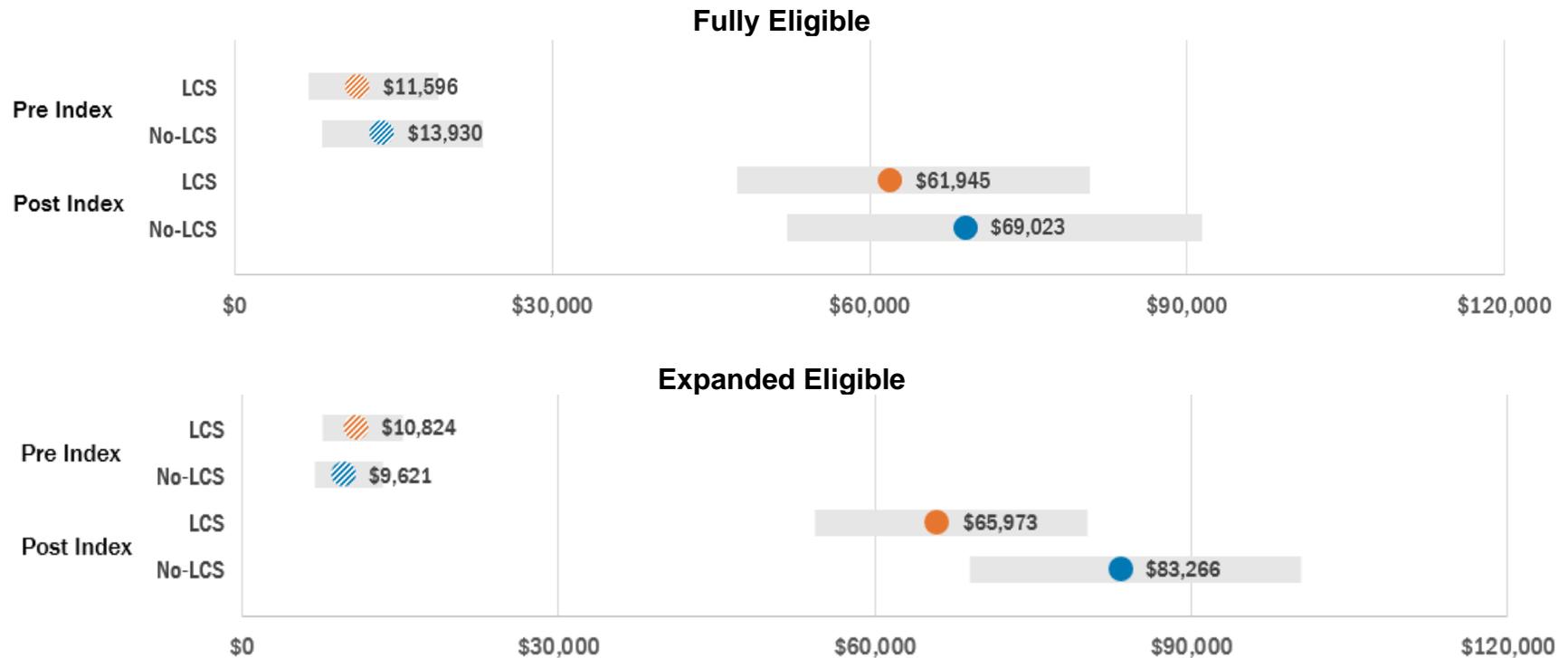

- Limited to lung cancer diagnoses occurring in year after index date.